\documentclass[preprint, pre]{revtex4-1}
\usepackage{graphicx} 
\usepackage{amsmath}
\usepackage{bm}
\usepackage{amssymb}
\usepackage{natbib}

\begin{document}
\title{Curvature Induced Activation of a Passive Tracer in an Active Bath}

\author{S. A. Mallory$^1$, C. Valeriani$^{2}$, A. Cacciuto$^1$}
\email{ac2822@columbia.edu}
\affiliation{$1$Department of Chemistry, Columbia University\\ 3000 Broadway, New York, NY 10027\\ }
\affiliation{$^{2}$Departamento de Quimica Fisica, Facultad de Ciencias Quimicas, Universidad Complutense de Madrid, 28040 Madrid, Spain}

\begin{abstract}
We use numerical simulations to study the motion of a large asymmetric tracer immersed in a low-density suspension of self-propelled particles in two dimensions. Specifically, we analyze how the curvature of the tracer affects its translational and rotational motion in an active environment. We find that even very small amounts of curvature are sufficient for the active bath to impart directed motion to the tracer, which results in its effective activation. We propose simple scaling arguments to characterize this induced activity in terms of the curvature of the tracer and the strength of the self-propelling force. Our results suggest new ways of controlling the transport properties of passive tracers in an active medium by carefully tailoring their geometry.

\end{abstract}

\maketitle
\textit{Introduction} --In recent years, the behavior and dynamics of microstructures and colloidal particles immersed in an active fluid (e.g. bacteria, self-propelled nanoparticles, artificial microswimmers, etc.) have drawn considerable interest. The inherently non-equilibrium driving forces and stochastic nature of an active fluid give rise to phenomenological behavior that is quite remarkable including anomalous diffusion \cite{chen2007, wu2000, mino2011, leptos2009, morozov, valeriani2011}, tunable effective interactions between suspended microcomponents \cite{angelani2011, harder2014, reichardt2014, ni2014}, and targeted delivery of colloids \cite{palacci2013, koumakis2013}. { In addition, there has recently been an effort to understand the dynamics, phase behavior, and self-assembly properties of suspensions of deformable \cite{ohta2009, menzel2012} and/or irregular shaped active particles \cite{wensink2014, nguyen2014}.} An emerging area in this field is designing microstructures to perform specific tasks when immersed in an active suspension, most notably driving microscopic gears and motors \cite{dileonardo2010, angelani2009}, the capture and rectification of active particles \cite{ghosh2013, wan2008, galajda2007, kaiser2012, kaiser2013}, and using active suspensions to propel wedge-like carriers \cite{angelani2010, kaiser2014}. The geometry of these microdevices is a crucial component to being able to effectively convert the energy from the active environment into mechanical work. 
Interestingly, Angelani and Di Leonardo \cite{angelani2010} showed that chevron shaped micro-shuttles immersed in a bacterial suspension undergo directed motion along their axis of symmetry. A similar observation was made experimentally by Kaiser et al. \cite{kaiser2014} who showed that chevron shaped particles can be set into rectified motion along their wedge cusp when immersed in a high density bacterial suspension. Given these results, it is well established that asymmetric tracers with locally concave regions (e.g. wedge, chevron, lock and key colloids, etc.) are able to undergo rectified motion in an active fluid, while spherical tracers only undergo enhanced isotropic diffusion \cite{wu2000, morozov, valeriani2011}. Intriguingly and in stark contrast { to a tracer in a passive environment}, the transport properties of a tracer immersed in an active fluid are strongly dependent on its underlying geometry. In short, a tracer in an active environment can be made active in its own right by simply altering its shape.

In an effort to characterize this unique phenomenon, we systematically distort the geometry of a rod shaped tracer and study the resulting dynamics in an active medium. Our goal is to understand the transition from isotropic to directed motion as a function of tracer geometry. In previous studies, the rectification of the random motion of the bacteria is caused by polar ordering and trapping of bacteria inside the cusp regions of the tracer. We however consider a low-density suspension of non-aligning active particles. 
This choice is motivated by the recent developments in the design and synthesis of artificial self-propelled particles \cite{Theurkauff2012, palacci2013, Buttinoni2, sacanna2}, as well as to eliminate the polar ordering and trapping which is typical of high density bacterial suspensions, {and be able to  focus exclusively on the effect of the particle activity disregarding collective effects that may ensue due to excluded volume interactions at larger concentrations.}
 Our results show that directed motion of the tracer can be obtained under a much more general set of geometric constraints and explain how it can be easily controlled by the curvature of the tracer alone. In other words, independently of the local ordering of the particles, induced activity can be imparted by local density gradients around the tracer, which can be tuned and enhanced by manipulating the curvature of the tracer.

\textit {Model} -- We consider a two dimensional model where a single asymmetric tracer is immersed in a bath of $N$ spherical active particles of diameter $\sigma$. Each active bath particle has mass $m$, and undergoes Langevin dynamics at a constant temperature $T$. Self-propulsion is introduced through a directional force of constant magnitude $|F_a|$ and is directed along a predefined orientation vector $\boldsymbol{n}=(\cos \theta, \sin \theta)$ which passes through the origin of each particle and connects its poles. The equations of motion of an individual particle are given by the coupled Langevin equations

\begin{align}
m \ddot{ \boldsymbol{r}} & =-\gamma \dot{ \boldsymbol{r}}+|F_a|{\boldsymbol n}-\partial_{\boldsymbol r} V+\sqrt{2\gamma^2D} {\boldsymbol \xi}(t) \\
\dot{\theta}& =\sqrt{2 D_r} \,\,\xi_r(t) 
\end{align}

\noindent where $\gamma$ is the translational friction and $V$ the interparticle potential acting on the particle The translational and rotational diffusion constants are given by $D$ and $D_r$, respectively. The typical solvent induced Gaussian white noise terms for both the translational and rotational motion are characterized by $\langle \xi_i(t) \rangle = 0$ and $\langle \xi_i(t) \cdot \xi_j(t') \rangle = \delta_{ij}\delta(t-t')$ and $\langle \xi_{r}(t)\rangle = 0$ and $\langle \xi_{r}(t) \cdot \xi_{r}(t') \rangle =\delta(t-t')$, respectively. { The translational diffusion constant $D$ is related to the temperature $T$ via the Stokes-Einstein relation $D=k_BT/\gamma$.} In the low Reynolds number regime, the rotational and translation diffusion coefficients for a sphere satisfy the relation $D_r=(3D)/\sigma^2$. 

\begin{figure}[h!]
\includegraphics{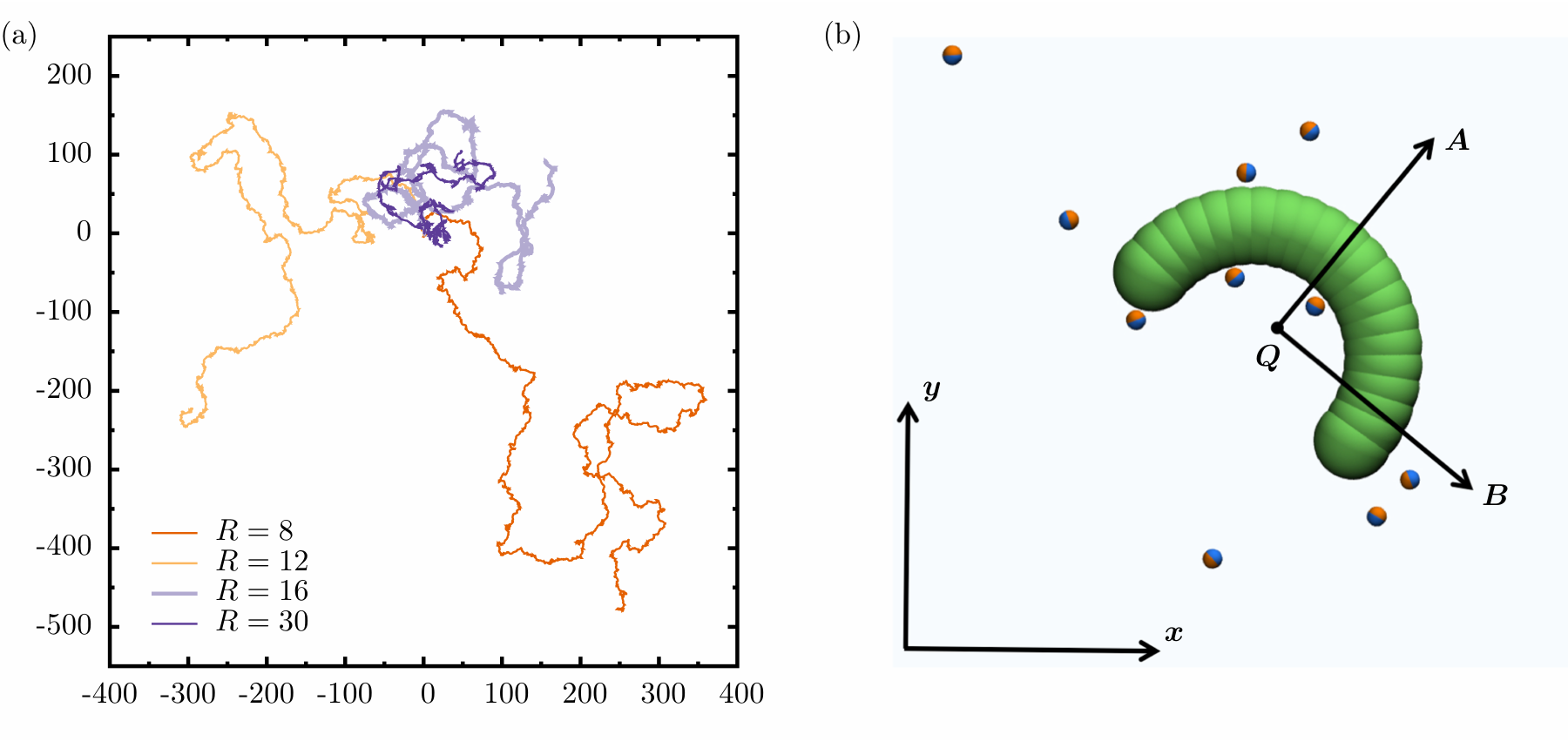}
\caption{\label{traj} {(Color online) (a) Trajectories for tracers of various curvatures immersed in an active bath of volume fraction $\Phi=0.005$, $T=1$, and $|F_a|=100$. The trajectory for each tracer was taken over a duration of $5000\tau$ and each tracer was initially located at the origin. (b) Snapshot from simulation with $R=8$, $\Phi=0.005$, $T=1$, and $|F_a|=100$ where both the laboratory and body-centered reference frames are shown. The orange half of a bath particle denotes the direction of propulsion.}}
\end{figure}

\begin{figure}[b!]
\includegraphics{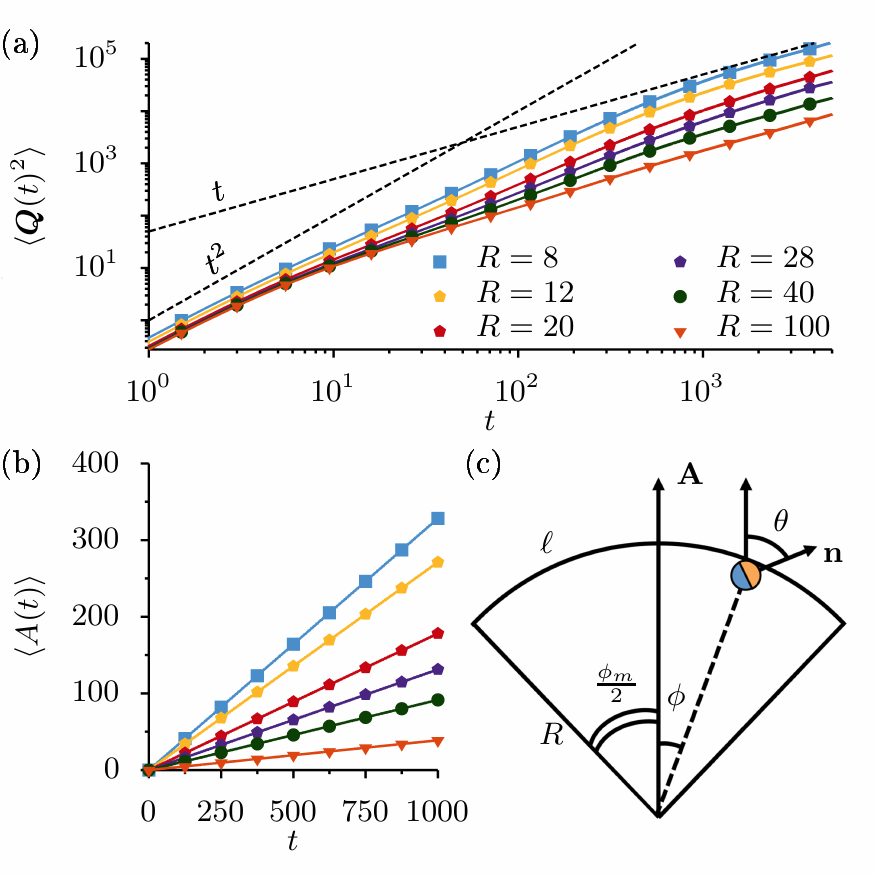}
\caption{\label{msd} {(a) MSD for tracers of various curvatures immersed in an active bath of volume fraction $\Phi=0.005$, $T=1$, and $|F_a|=100$. (b) Mean displacement of the tracer along the $A$ axis in the body centered frame. (c) Sketch of the 
model and relative variables discussed in the paper. }}
\end{figure}

Each tracer is modeled as a bent rod characterized by its arc length $\ell=25\sigma$ and radius of curvature $R$ (Fig. \ref{traj}(b) and Fig. \ref{msd}(c)). For practical purposes, the rods are discretized into $N_T=21$ equidistant and overlapping spherical particles of diameter $\sigma_T=2.5\sigma$. A suitably large number of spheres were chosen to accurately reproduce the shape of the particle and to make the surface sufficiently smooth. The tracer itself undergoes over-damped Langevin dynamics at a constant temperature $T$ and the equations of motions are the rigid body analogs to Eqs. (1) and (2) where $|F_a|=0$ since the tracer itself is non-active. 

All interactions between the particles in the systems are purely repulsive and are given by the Weeks-Chandler-Andersen (WCA) potential
\begin{equation}
U(r_{ij})=4 \epsilon \left[ \left( \frac{\sigma_{ij}}{r_{ij}} \right)^{12}- \left( \frac{\sigma_{ij}}{r_{ij}} \right)^{6}+\frac{1}{4} \right] 
\label{eq:LJ_V}
\end{equation} 
\noindent with a range of action extending up to $r_{ij}=2^{1/6}\sigma$. Here $r_{ij}$ is the center to center distance between any two particles $i$ and $j$, $\sigma_{ij}=(\sigma_i+\sigma_j)/2$ where $i=1,2$ corresponding to an active particle or a tracer particle, respectively, and $\epsilon=10k_{\rm B}T$ is the interaction energy. { Using the numerical package LAMMPS ~\cite{plimpton1995}, all simulations were carried out in a periodic box of dimension $L=200$ with $T=m=\sigma=\tau=1$ and $\gamma=10\tau^{-1}$(here $\tau$ is the dimensionless time). Each simulation was run for a minimum of $5 \times10^8 \tau$ time steps. {The drag coefficient $\gamma$ was chosen to be sufficiently large such that the motion of the particles is effectively overdamped. Several of the simulations were repeated with larger values of $\gamma$ (e.g. $\gamma=50,100$), which produced no detectable differences in our results.} All quantities in this investigation are given in reduced Lennard-Jones units.} 

\textit{Results} -- {As a way of illustrating our main result we show in Fig. 1(a) typical trajectories (particle traces) for tracers having different radii of curvature $R$ at constant arc length $\ell$ immersed in an active suspension of volume fraction $\Phi =0.005$ and propelling force $|F_a|=100$. A sufficiently large propelling force was chosen to illustrate the curvature induced activation of a tracer. In addition, a typical snapshot from a simulation is given in Fig. \ref{traj}(b), which details the different reference frames used in the subsequent analysis.} The typical mean square displacement (MSD) of the center of mass $\boldsymbol{Q}$ for various tracers in the laboratory frame is shown in Fig.~\ref{msd}(a). All tracers undergo ballistic behavior at short times, and eventually crossover to a diffusive regime at longer times. As the curvature of the tracer increases, the super-diffusive regime persists for longer times { prolonging the eventual crossover to the purely diffusive regime}. In the body-centered frame, the mean displacement of the tracer along its main axis of symmetry $\boldsymbol{A}$ is given in Fig.~\ref{msd}(b). For a straight tracer (i.e. $R=\infty$), there is no net displacement along the main axis of the tracer, which is obvious from symmetry considerations. However, as soon as the symmetry of the tracer is broken by introducing any amount of curvature, the tracer undergoes net directed motion in the positive $\boldsymbol{A}$ direction leading to a nonzero mean displacement in the body-centered frame. As the tracer becomes increasingly curved (Fig. \ref{msd}(b)), this effective propelling force becomes larger. 

\begin{figure}[h!]
\includegraphics{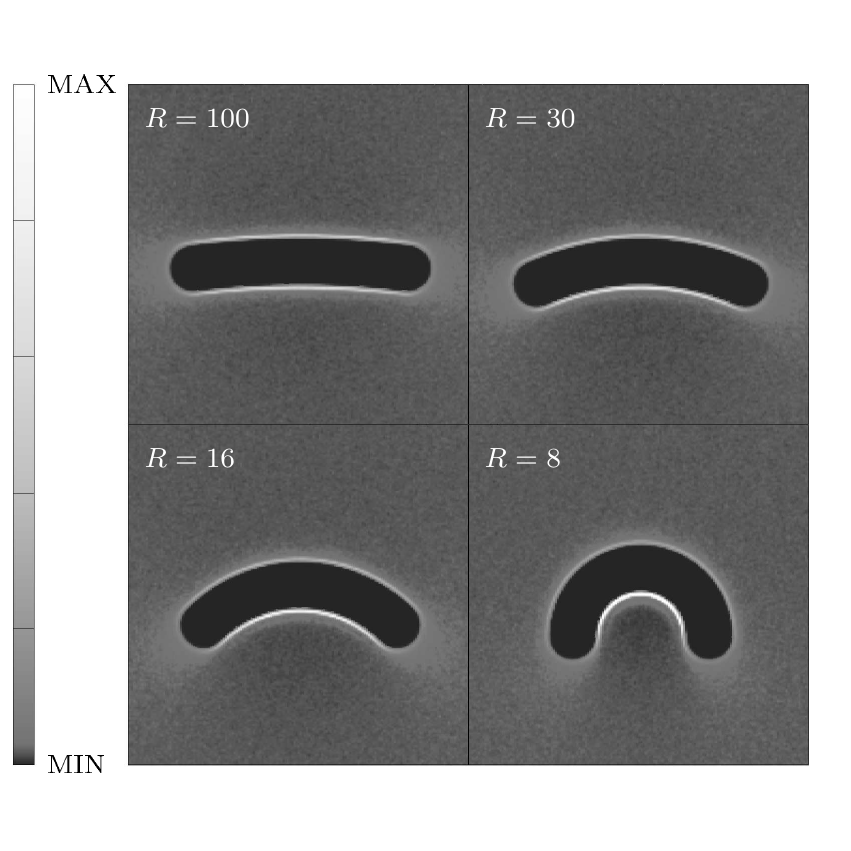}
\caption{\label{denmap} {Normalized time averaged active particle density for various tracers in an active suspension of $\Phi=0.005$, $T=1$, and $|F_a|=100$. }}
\end{figure}

The directed motion of the tracer can be understood by looking at the { time averaged} local density of active particles in the system (Fig. \ref{denmap}). The average local density is homogeneous across the system, except along the surface of the tracer, where it shows a significant increase. Specifically, the density is larger on the concave side of the tracer when compared to its convex side, and this difference becomes even larger for tracers with higher degrees of curvature. As demonstrated in reference~\cite{fily2014}, the positive curvature of a surface can act as a restoring force against random thermal rotations, and drives active particles towards a stable orientation where the propelling axis becomes parallel to the surface normal. This stabilizing effect greatly increases the time required for a particle to escape from the side of the surface with positive curvature. The side of the tracer with negative curvature behaves in the opposite way and destabilizes the axial angle of the particle from the surface normal upon any amount of thermal rotations, which results in a significantly shorter escape time. The combination of these two mechanisms produces the measured density gradient across the tracer that results in its net directed motion.

\begin{figure}[ht!]
\includegraphics{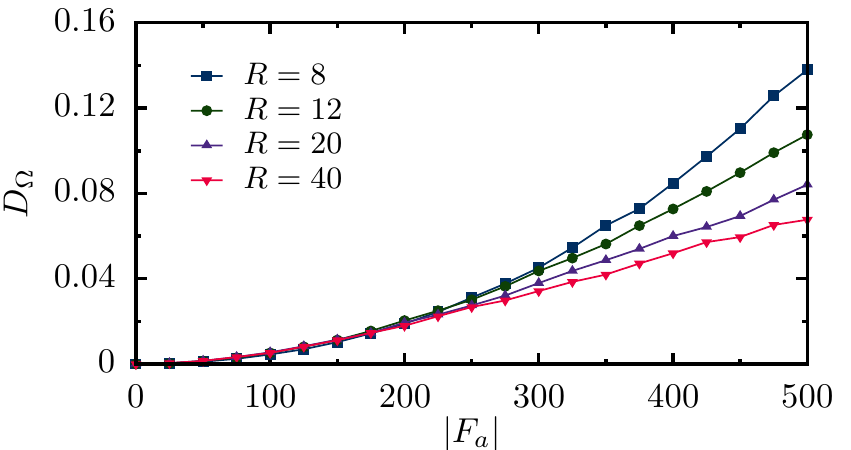}
\caption{\label{angdiff} {Rotational diffusion constants for tracer particles having different curvatures in an active suspension of $\phi=0.005$ and $T=1$. The data is plotted as a function of the active bath self-propelling force $|F_a|$.}}
\end{figure}

Due to the symmetry of the system, the average tangential force as well as the average torque will be equal to zero, and indeed we find that the mean squared angular displacement of the tracer $\langle \Omega^2 \rangle$ is diffusive for almost all observed times (i.e. $\langle \Omega^2 \rangle = D_\Omega t$ where $D_\Omega$ is the rotational diffusion constant of the tracer). We find that the rotational diffusion constant $D_\Omega$ increases with the strength of the active force as well as with the curvature of the tracer for large self-propelling forces (Fig. \ref{angdiff}). { We however defer a full characterization of the rotational dynamics to a later publication, as it is a highly non-trivial problem that truly deserves its own in-depth analysis. The remainder of this work focuses on developing a scaling theory for the curvature induced activation of the tracer.}

\textit{Discussion} -- The directed motion of the curved tracer emerges when the persistent length of an active particle becomes much larger than the dimension of the tracer. In this regime, a simple way of estimating the net force exerted on the tracer can be obtained by considering that a bath particle can { either be on the surface of the tracer pushing with a force proportional to $|F_a|$} or diffusing across the system and applying no force. As previously mentioned, the amount of time an active particle spends on the surface of the tracer is highly dependent on the side it is located and characterized by a residence time $\tau_n$ or $\tau_p$ for the side of the tracer with negative or positive curvature, respectively. { Given these quantities and the { average time an active particle spends in the bath between collision with the tracer}, which we denote by $\tau_0$, the net average force exerted by an ideal gas of $N$ active particles} along the $\boldsymbol{A}$ axis of the tracer can be estimated as $\langle F_{A}\rangle = N(\langle F_p\rangle-\langle F_n\rangle)$, where $\langle F_p\rangle$ and $\langle F_n\rangle$ are the average forces exerted on the positive and negative sides of the tracer. In the low density and small tracer limit (i.e. $\tau_0\gg \tau_n$ and $\tau_0\gg\tau_p$),

\begin{equation}
\langle F_{A} \rangle\simeq N \langle F\rangle \left (\frac{\tau_p -\tau_n} {\tau_0}\right )
\label{main}
\end{equation}

\noindent where $\langle F\rangle \simeq |F_a|\langle \cos(\phi)\rangle$ is the average force an active particle exerts along the $\boldsymbol{A}$ axis of the tracer. { For simplicity, we assume that the particle axis is predominantly parallel to the surface normal at that point~\cite{fily2014}. }

Given the geometry of the system and assuming that an active particle can diffuse anywhere on the surface of the tracer with equal probability (where the angular range spans $[-\frac{\phi_m}{2},\frac{\phi_m}{2}]$ (Fig. \ref{msd}(c)), it follows than an approximation for $\langle F \rangle$ is given by

\begin{equation}
\langle F\rangle= \frac{2|F_a|}{\phi_m}\sin(\frac{\phi_m}{2})
\end{equation}

In order to estimate $\tau_p$ and $\tau_n$, we use a similar approach to that proposed by Fily et al. \cite{fily2014}. In the limit of large activity, it is fair to assume that the active particles can only leave the surface of the tracer (once they are in contact with it) by sliding out. {This sliding motion is driven by the angle between the self-propelling axis and the boundary which results in tangential forces along the tracer}, and can be described with the coupled overdamped Langevin equations
\begin{align}
\dot{\phi} &= \frac{v_a}{R} \sin(\theta-\phi)\\
\dot{\theta}& =\sqrt{2 D_r} \,\,\xi_r(t) 
\end{align}
where $\theta$ is the angle of the active axis, and $\phi$ is the angular position of the active particle with respect to the osculating circle of the tracer (see Fig. \ref{msd}(c)) . Here $v_a=|F_a|/\gamma$ is the particle active velocity, and for simplicity, we neglected the thermal noise in $\phi$ as for large forces it gives only a small contribution to the sliding motion of the particle. 
We define $\alpha \equiv \theta-\phi$ to be the angle between the self-propelling axis and the boundary normal.
By taking the difference between the two equations, we have

\begin{equation}
\dot{\alpha}=-\frac{v_a}{R} \sin(\alpha) + \sqrt{2 D_r}\xi_r(t) \simeq -\frac{v_a}{R} \alpha+ \sqrt{2 D_r}\xi_r(t)
\label{alpha}
\end{equation}
for a particle facing the concave side of the tracer. { In the large activity regime, Fily et al. \cite{fily2014} have demonstrated that the small angle approximation in Eq. \eqref{alpha} is valid for sufficiently curved surfaces.} The equation for a particle on the convex side is obtained by replacing $R\rightarrow -R$.

This equation can be readily solved to give
\begin{equation}
\langle \alpha^2(t) \rangle = \frac{RD_r}{v_a}\left(1- e^{-2\frac{v_a}{R} t} \right)
\end{equation}
{ Notice that in this derivation we ignored the contribution due to $\alpha(0)$ (the initial angle of impact of a particle
with the tracer). This is justified because the time spent on the tracer by an active particle with a large value of $\alpha(0)$ is typically small and so is its contribution to the net force on the tracer. Furthermore, we find that the probability that a particle hits the tracer with a large angle with respect to the surface normal is exponentially small (data not shown). 
Nevertheless the contribution from $\alpha(0)$ becomes critical in the limit of very short tracers, when the active particles slide off before any significant rotational diffusion can take place.}

Since $\langle \alpha^2 \rangle= \langle \phi^2 \rangle+\langle \theta^2 \rangle-2\langle \phi \theta \rangle= \langle \phi^2\rangle-\langle \theta^2 \rangle+2\langle \alpha \theta \rangle $ where $\langle \theta^2 \rangle=2D_r t$ and 
\begin{equation}
\langle \alpha \theta \rangle = 2D_r\int^t_0 dt_1\int^t_0 dt_2 \langle \xi(t_1) \xi(t_2) \rangle e^{-\frac{v_a}{R}(t-t_1)}=\frac{2RD_r}{v_a}(1-e^{-\frac{v_a}{R}t}),
\end{equation} we can solve explicitly for $\langle \phi^2 \rangle$
\begin{equation}
\langle \phi^2 \rangle_p = 2D_rt+ \frac{RD_r}{v_a} \left[ \left(1-e^{-2\frac{v_a}{R}t}\right)-4\left(1-e^{-\frac{v_a}{R}t}\right) \right]
\label{phi2}
\end{equation}

\noindent Equation \eqref{phi2} gives the MSD of an active particle along the concave surface of the tracer (positive curvature). The analogous equation for the convex side of the tracer (negative curvature) is again obtained by replacing $R\rightarrow -R$, and gives
\begin{equation}
\langle \phi^2 \rangle_n = 2D_rt+ \frac{RD_r}{v_a} \left[ \left(e^{2\frac{v_a}{R}t}-1\right)-4\left(e^{\frac{v_a}{R}t}-1\right) \right]
\label{phi3}
\end{equation}

In the long time limit ($t \rightarrow \infty$), the particles facing the concave side of the traces will undergo standard diffusive behavior 
$\langle\phi^2\rangle_p=2D_r t$, whereas those facing the convex side will have an exponentially growing angular dependence.

{ The expression for the MSD on the concave side of the tracer (Eq. \eqref{phi2}) does indeed confirm that the positive curvature acts as a restoring force which greatly increases the time required for a bath particle to escape, while (Eq. \eqref{phi3}) clearly reveals that the side of the tracer with negative curvature destabilizes the propelling axis of an active particle away from the surface normal giving rise to a faster escape time.} Expanding Eqs. \eqref{phi2} and \eqref{phi3} for small times, we obtain
\begin{equation}
\langle \phi^2 \rangle_{p/n}\simeq\frac{2}{3}D_r\left(\frac{v_a}{R}\right)^2 t^3
\label{qqq}
\end{equation}
For the case of particles on the concave surface, we verified these results numerically by explicitly measuring the diffusion of an active particle confined inside a circular cavity.

\begin{figure}[h!]
\includegraphics{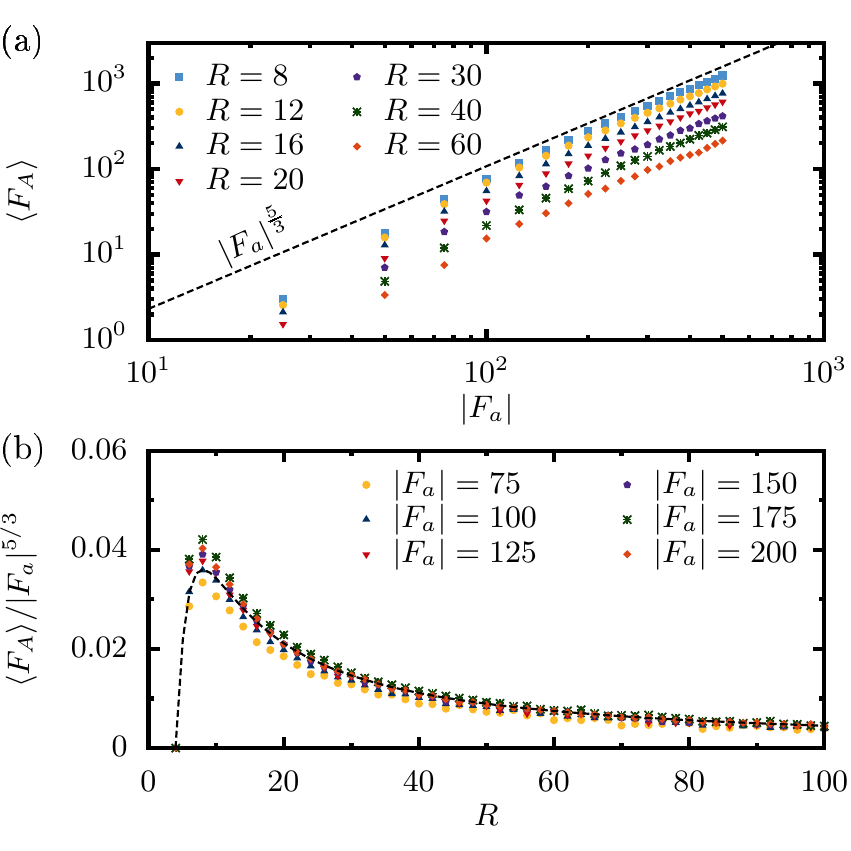}
\caption{\label{force} {(a) Effective force, $\langle F_A\rangle$, induced by the active particle on the tracer as a function of propulsion $|F_a|$ for various tracer curvatures. (b) Collapse of $\langle F_A\rangle$ as a function of $R$ for different values of $|F_a|$. { The dashed line corresponds to Eq. \eqref{final} with a single fitting parameter $\kappa=0.40(0)$ in the prefactor. }}}
\end{figure}

We define $\tau_{p}$ and $\tau_{n}$ to be the times required for the particles facing respectively
the concave and the convex sides of the tracer to cover the same arc length 
$\langle \phi^2 \rangle_{p}=\langle \phi^2 \rangle_{n}=\left(\frac{\ell}{R}\right)^2$. 
We first discuss the limit of large $R$ or small $\ell$, for which $\tau_{p}\gtrsim\tau_{n}$.
Carrying out the expansion of Eqs.~\eqref{phi2} and \eqref{phi3} to the fourth order
and taking the difference of the two expressions gives 
\begin{equation}
\frac{2}{3}\left(\frac{v_a}{R}\right)^2(\tau_p^3-\tau_n^3)=\frac{1}{2}\left(\frac{v_a}{R}\right)^3(\tau_p^4+\tau_n^4)
\label{lll} 
\end{equation}
For sufficiently large $R$, we can write $\tau_{p}-\tau_{n}=\varepsilon$, where $\varepsilon$ 
is a small but positive number. We can then further expand Eq.~\eqref{lll} in $\varepsilon$ by writing
$\tau_{p}^3-\tau_{n}^3\simeq 3\tau_{n}^2\varepsilon$ and $\tau_p^4+\tau_n^4\simeq 2\tau_n^4$, to get
$\varepsilon=\frac{1}{2}\left ( \frac{v_a}{R}\right ) t_n^2$. Using Eq.~\eqref{qqq}, with 
$\langle \phi^2 \rangle=\left(\frac{\ell}{R}\right)^2$, we finally get $\tau_n^2=\left(\frac{3\ell^2}{2D_rv_a^2}\right)^{\frac{2}{3}}$,
and thus $\varepsilon=\tau_p-\tau_n=\frac{1}{2R}\left(\frac{3}{2D_r}\right)^{\frac{2}{3}}\ell^{\frac{4}{3}} v_a^{-\frac{1}{3}}$.

{ The average collision rate between a single bath particle and the tracer can be estimated by $1/\tau_0=(1/L^2) C v_a$ where $C$ is the collision cross section of the tracer.} A reasonable approximation for the collision cross section is either the length of the tracer $l$ for relatively straight tracers or the radius of curvature $R$ for highly curved tracers. 

Using Eq.~\eqref{main}, and taking $1/\tau_0\simeq{(1/L^2)\ell v_a}$, we finally obtain
\begin{equation}
\langle F_A \rangle \simeq \frac{\rho\ell^{\frac{4}{3}} |F_a|^{\frac{5}{3}}}{(D_r\gamma)^{\frac{2}{3}}} \sin\left(\frac{\ell}{2R}\right)
\label{final}
\end{equation}
where $\rho=N/L^2$. We find that this functional form properly accounts for all of our data. This is shown in Fig. \ref{force} where Eq.~\eqref{final} has been used
to fit the data both in terms of the radius of curvature $R$ and the strength of activity $|F_a|$ { with a single fitting parameter in the prefactor given by $\kappa=0.40(0)$}.

We expect deviations from this law to appear for long and highly curved tracers, where in general a significant amount of diffusive sliding occurs before the particles leave the
tracer. In this case, {the small $R$ limit} of Eqs.~\eqref{phi2} and ~\eqref{phi3} is more appropriate, and $\tau_p$ can be approximately written as $\tau_p\simeq \frac{1}{2D_r} \left(\frac{\ell}{R}\right)^2$. In this limit $\tau_p\gg\tau_n$ and $1/\tau_0$ can be written as $1/\tau_0\simeq (1/L^2) Rv_a$ resulting in 
\begin{equation}
\langle F_A \rangle \simeq \frac{\rho\ell |F_a|^2}{\gamma D_r}\sin\left(\frac{\ell}{2R}\right)
\end{equation}
It is important to stress that in both cases we have the same curvature dependence of the effective active force,
and that curvature is the crucial parameter for the activation of the tracer with a dependence given by $\langle F_A\rangle\sim 1/R$.

\textit {Conclusions}-- Using a combination of numerical simulations and analytical theory, we have demonstrated how a tracer can be made effectively active when immersed in a suspension of active particles. We have analyzed how the speed of this effective motion can be enhanced with the curvature of the tracer, and proposed simple theoretical arguments to quantify the induced activity as a function of the strength of the bath activity and the tracer curvature. Our results are most valid in the low density limit, where the residence time of the active particles on the surface of the tracer is much smaller than the typical time required for the particles in the bulk to find the tracer. Clearly whenever, the two timescales are of the same order a crossover from a super-linear to a linear dependence of the effective force on $|F_a|$ should be expected, at least in the ideal gas limit. In fact, at higher densities, when significant clustering on the concave side of the surface occurs, excluded volume interactions become important and will act to weaken the overall force exerted on the wall as explained in our previous work~\cite{mallory2014}. Indeed, the spherical shape of the active particles will not produce any cooperative alignment (at least as long as hydrodynamic interactions are not considered), but will prevent optimal ordering of the particles propelling axes along the normal to the surface.

{ In principle, one could improve our estimates for the calculation of $\langle F_A\rangle$ by modifying 
$\langle F\rangle$ in Eq. (5) to also include the average over the small deviations of the active particle axis away from the normal to the surface. This gives
\begin{equation*}
\langle F\rangle =2|F_a| \frac{\sin(\phi_m/2)}{\phi_m} \exp\left({-\frac{\gamma RD_r}{|F_a|}}\right)
\end{equation*} 
but for large forces and sufficiently curved tracers, this would only add a sub-leading term to our estimates. }

It should be finally noted that in all our simulations we considered the friction coefficient of the tracer $\gamma_t$ to be independent of the particle curvature. Clearly this is an approximation as we expect this value to be dependent on $R$. Unfortunately, evaluating an explicit formula for $\gamma_t (R)$ is not trivial, but any curvature dependence could be easily incorporated. 

\textit {Acknowledgments}--We thank Clarion Tung and Joseph Harder for insightful discussions and helpful comments. AC acknowledges financial supported from the National Science Foundation under CAREER Grant No. DMR-0846426. CV acknowledges financial support from a Juan de la Cierva Fellowship, from the Marie Curie Integration Grant PCIG-GA-2011-303941 ANISOKINEQ, and from the National Project FIS2013-43209-P. SAM acknowledges financial support from the National Science Foundation Graduate Research Fellowship. This work used the Extreme Science and Engineering Discovery Environment (XSEDE), which is supported by National Science Foundation grant number ACI-1053575.

\bibliographystyle{apsrev4-1}

\bibliography{passive2active}

\end{document}